\documentclass[useAMS,usenatbib]{mn2e}
\usepackage{graphicx}
\usepackage{epstopdf}
\usepackage{subfigure}
\usepackage{lscape}
\usepackage{natbib}
\usepackage{amssymb,bm}
\usepackage{epsfig}
\usepackage{rotating}
\usepackage{fancyhdr}
\usepackage{mathrsfs}
\usepackage{amsmath}
\usepackage{amssymb}
\usepackage{verbatim}
\title[kHz QPOs in 4U 1636--53]{The time derivative of the kilohertz quasi-periodic oscillations in 4U 1636--53 }
\author[Sanna et al.]{Andrea Sanna$^{1}$\thanks{E-mail: A.Sanna@astro.rug.nl}, Mariano M\'endez$^{1}$, Tomaso Belloni$^{2}$, Diego Altamirano$^{3}$\\
$^{1}$Kapteyn Astronomical Institute, University of Groningen, P.O. BOX 800, 9700 AV Groningen, The Netherlands\\
$^{2}$INAF-Osservatorio Astronomico di Brera, Via E. Bianchi 46, I-23807 Merate (LC), Italy\\
$^{3}$Sterrenkundig Instituut Anton Pannekoek, Science Park 904, 1098 XH Amsterdam, the Netherlands}

\date{Accepted 2012 May 31.  Received 2012 May 24; in original form 2012 March 21
      }

\pagerange{\pageref{firstpage}--\pageref{lastpage}} \pubyear{2002}

\begin{document}

\label{firstpage}
\maketitle
\begin{abstract}
We analysed all archival RXTE observations of the neutron-star low-mass X-ray binary 4U 1636--53 up to May 2010. In 528 out of 1280 observations we detected kilohertz quasi-periodic oscillations (kHz QPOs), with $\sim 65$\% of these detections corresponding to the so-called lower kHz QPO. Using this QPO we measured, for the first time, the rate at which the QPO frequency changes as a function of QPO frequency. For this we used the spread of the QPO frequency over groups of 10 consecutive measurements, sampling timescales between 320 and 1600 s, and the time derivative of the QPO frequency, $\dot \nu_{\rm QPO}$, over timescales of 32 to 160 s. We found that: (i) Both the QPO-frequency spread and $\dot \nu_{\rm QPO}$ decrease by a factor $\sim 3$ as the QPO frequency increases. (ii) The average value of $\dot \nu_{\rm QPO}$ decreases by a factor of $\sim 2$ as the timescale over which the derivative is measured increases from less than 64 s to 160 s. (iii) The relation between the absolute value of $\dot \nu_{\rm QPO}$ and the QPO frequency is consistent with being the same both for the positive and negative QPO-frequency derivative.
We show that, if either the lower or the upper kHz QPO reflects the Keplerian frequency at the inner edge of the accretion disc, these results support a scenario in which the inner part of the accretion disc is truncated at a radius that is set by the combined effect of viscosity and radiation drag.
\end{abstract}


\section{Introduction}
Kilohertz quasi-periodic oscillations (kHz QPOs) are the fastest variability so far observed in neutron star (NS) low-mass X-ray binary (LMXB) systems. These oscillations were first detected (\citealt{v1996a}; \citealt{st1996}) shortly after the launch of the Rossi X-ray Timing Explorer (hereafter RXTE; see \citealt{b1993}; \citealt{ja2006}). Since then, kHz QPOs have been observed in more than 30 neutron star LMXBs (see, e.g.,  \citealt{v2005}). KHz QPOs are often observed in pairs, usually called lower and upper kHz QPO, with frequencies ranging from a few hundred Hz to more than 1 kilohertz (e.g., \citealt{v2005}, for an extensive review). The fact that the timescale of these oscillations corresponds to the dynamical timescale very close to the NS, makes kHz QPOs potentially one of the few tools nowadays available to directly measure strong-field gravitational effects.

In the past 16 years, several models have been proposed to explain these oscillations, with emphasis in trying to explain the frequency of the QPOs. Almost all models associate the frequency of these oscillations with characteristic frequencies in a geometrically thin accretion disc (e.g., \citealt{mlp1998}; \citealt{sv1998}). Systematic studies of other properties of kHz QPOs, such as the coherence and the amplitude of the oscillation, are also available for several LMXBs (\citealt{j2000}; \citealt{van2000}; \citealt{ds2001}; \citealt{mvf2001}; \citealt{h2002}; \citealt{van2002}; \citealt{dmv2003}; \citealt{van2003}; \citealt{b2005a}; \citealt{b2005b}; \citealt{bom2006}; \citealt{a2008}; \citealt{sa2010}), but not many models explain these other properties (see \citealt*{mlp1998}; \citealt*{bom2006}). Some sources show a characteristic dependence of the coherence (defined as $Q = \nu/\Delta \nu$, where $\nu$ and $\Delta\nu$ are the centroid frequency and the full width at half-maximum of the kHz QPO, respectively) and the amplitude of those oscillations upon the centroid frequency of the oscillation: Both the coherence and the amplitude of the lower kHz QPO in these sources increase as the frequency increases, and then quickly decrease as the QPO frequency continues to increase. \citet*{bom2006} proposed that the drop of the coherence $Q$ at high frequencies is a direct consequence of the disc approaching the innermost circular stable orbit (ISCO), where the inner edge of the disc gets shattered by the gravitational force of the neutron star. If this interpretation is correct, the drop of the coherence and rms amplitude of the lower kHz QPO at high frequencies would confirm the existence of the ISCO, and it would provide a direct measurement of the mass of the neutron star \citep*{mlp1998}. However, other non space-time-related interpretations have been proposed to explain the properties of QPOs; e.g., \citet{m2006} suggested that the drop of $Q$ and rms amplitude in individual sources might be related (at least in part) to changes in the properties of the accretion flow in these systems. This could explain the significantly different properties of the kHz QPOs, respectively, in the high and low-luminosity phase of the outburst of the LMXB XTE J1701-462 \citep{sa2010}.

Here we investigate the properties of the kHz QPOs for the NS system 4U 1636-53 by scanning the whole RXTE archive until May 2010. In Section 2 we describe the observations and the data analysis, and in Section 3 we present our results. In Section 4 we discuss our results in the context of current ideas concerning the origin of the kHz QPOs in LMXBs.

\section[]{OBSERVATIONS AND DATA ANALYSIS}
\label{data}

We analysed all archival observations of the LMXB 4U 1636--53 obtained with the Proportional Counter Array (PCA) on board of RXTE up to May 2010. We used a total of 1280 RXTE observations, which correspond to an exposure time of $\sim$ 3.5 Ms. During these observations the source showed a large number of type-I X-ray bursts that we excluded from our analysis (see, \citealt{z2011}).

Except for four observations, we used event-mode data with 125$\mu s$ time resolution and 64 channels covering the full PCA energy band to produce Fourier power density spectra (PDS), setting the time resolution to 4096 points per second, corresponding to a Nyquist frequency of 2048 Hz. In the four observations in which the event-mode data were not available, we used a combination of binned modes that allowed us to reach the same Nyquist frequency in the PDS. In all cases we selected photons with energies below $\sim$46 keV (absolute channel 101), with the exception of observation 10072-03-01-00 where we selected photons between channels 24 and 139, and observations 10072-03-02-00 and 10088-01-08-010 where we used the full energy band, because in these cases it was not possible to have simultaneously the same Nyquist frequency and energy selection as in the other observations. We created Leahy-normalised PDS for each 16-seconds data segment. We removed detector drop-outs, but no background subtraction or dead-time correction were applied before calculating the PDS. \\
We averaged all 16-s spectra within each observation, and we searched for kHz QPOs by fitting the averaged PDS in the frequency range 200-1500 Hz using a constant to model the Poisson noise and one or two Lorentzians to model the QPOs. From each fit, we estimated the significance of the Lorentzian by dividing its normalisation by the negative $1\sigma$ error. We considered only features with this ratio larger then 3, and coherence Q larger than 2.\\
We calculated the X-ray colours and intensity of the source using the Standard 2 data (16-s time resolution and 129 channels covering the entire PCA energy band). We defined soft and hard colour as the count rate ratio in the energy bands 3.5-6.0 keV / 2.0-3.5 keV and 9.7-16.0 keV / 6.0-9.7 keV, respectively. The intensity of the source is defined as the count rate in the energy band 2.0-16.0 keV. To obtain the exact count rate in each of these energy bands we linearly interpolated in channel space, since the energy boundaries of each channel change slightly with time. To correct for gain changes and differences in the effective area between the proportional counter units (PCUs) as well as differences due to changes in the channel to energy conversion of the PCUs as a function of time, we normalised the colours and intensity to those of the Crab Nebula obtained close in time to our observations (see \citealt{k2004} and \citealt{a2008} for details). Finally we averaged the normalised colours and intensities per PCU for the full observation using all available PCUs.\\

\subsection{KHz QPOs Identification}
\label{QPOiden}
\begin{figure}
\begin{center}
\includegraphics[width=80mm]{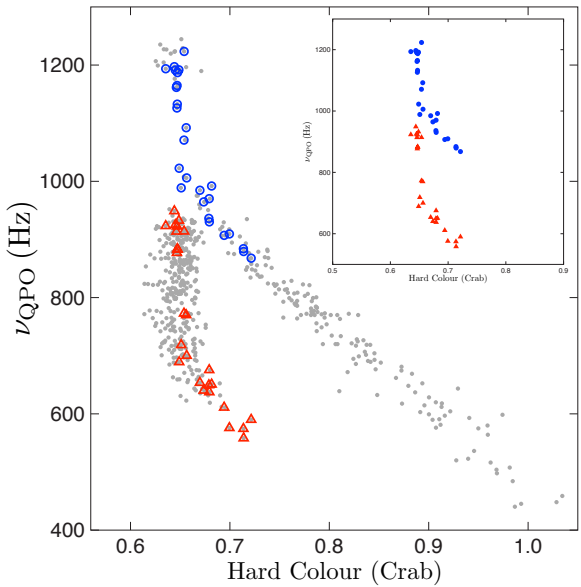}
\caption{Centroid frequency of the kHz QPOs detected in 4U 1636-53 as a function of hard colour. Grey bullets represent the QPOs in observations in which only one kHz QPO is observed. Red triangles and blue circles represent QPOs identified, respectively, as the lower and the upper kHz QPO in observations in which two simultaneous kHz QPOs are detected. For clarity, in the inset we plot only the QPO frequencies of the observations in which two simultaneous kHz QPOs were detected.}
\label{hard_freq}
\end{center}
\end{figure}

We detected kHz QPOs in 528 out of 1280 observations analysed. Among the observed QPOs, $\sim$84\% where detected with a significance larger then 3.5$\sigma$, $\sim$72\% with a significance larger then 4$\sigma$, and $\sim$53\% with a significance larger then 5$\sigma$. Even though $\sim$16\% of the QPOs could be considered only marginally significant (between 3 and 3.5$\sigma$), all these QPOs follow the frequency-hardness correlation described below, which adds confidence to the detections. Only 26 observations with QPOs showed two simultaneous high-frequency oscillations. In those cases the QPO identification is trivial because lower and upper kHz QPOs were both present. For all the other observations where we only detected a single QPO peak we had to apply a different method to identify the peak as either the lower or the upper kHz QPO. In Figure \ref{hard_freq} we plot, for each observation, the centroid frequency of all the kHz QPOs as a function of the hard colour of the source. From the plot it is apparent that, for frequencies below 1000 Hz, the data follow two separate branches. The first branch extends between QPO frequencies of $\sim$ 550 Hz and $\sim$ 950 Hz, with more or less constant hard colour at around 0.65, except for a small deviation at low frequencies where the hard colour increases slightly. The second branch shows a clear anti-correlation between frequency and hard colour. This has been already reported for this source (see \citealt{b2007}) as well as for other neutron-star systems (e.g., \citealt{m1999}; \citealt{mv1999}) and can be used to identify the two different kHz QPOs: Lower kHz QPO at low hard colour, and upper kHz QPO at high hard colour.

In Figure \ref{hard_freq} we also show the kHz QPOs detected simultaneously per observation: Red-empty triangles and blue-empty circles represent, respectively, the lower and the upper kHz QPOs. Those points confirm the correlation previously mentioned between frequency, hard colour and type of QPO. Interestingly, by checking the distribution of the blue circles in Figure \ref{hard_freq} we can resolve the ambiguity regarding the identification of the QPOs in the region above 1000 Hz, in which the two branches approach each other. To summarise, we detected a total of 357 lower kHz QPOs and 197 upper kHz QPOs. Compared to \citealt{b2007}, we find that the branch corresponding to the upper kHz QPO extends to higher frequencies at more or less constant hard colour, while the branch corresponding to the lower kHz QPO extends to lower frequencies at higher hard colour.

\subsection{Variations of the kHz QPO frequency}
\label{sp}
It is well known that the frequency of the kHz QPOs can change over tens of Hz in time intervals of a few hundred seconds \citep[e.g.,][]{b1996}, and this can artificially broaden the QPO in the averaged power spectrum of long observations. In order to study frequency variations of the kHz QPOs, we tried to detect kHz QPOs on timescales shorter than the full observation. We found that in observations with only the upper kHz QPO we cannot detect  the QPO significantly on timescales shorter than the full observation, whereas we detected the lower kHz QPO on timescales between 32 s and 160 s in about 50\% of the observations that showed kHz QPOs, all at significance levels larger than 4$\sigma$. For these observations we created dynamical power density spectra \citep[see, e.g., Fig.2 in][]{b1996}, and we extracted the frequency evolution as a function of time (hereafter frequency profile). To reduce the influence of the frequency error on the analysis of the frequency variation, we applied a Savitzky--Golay filter (\citealt{sk1964}; \citealt{nr}) to smooth the frequency profile. This method locally fits a polynomial function on a series of values (in our case we used a 4th order polynomial on 6\footnote{We varied the number of overlapping points from 4 to 10, but within errors this does not affect the time derivative of the QPO frequency.} consecutive points), and it provides the first time derivative of the QPO frequency (hereafter frequency derivative) for each measurement.

For each observation we then calculated how much the centroid frequency moved with time by measuring the sample standard deviation (hereafter the spread) using groups of 10 consecutive measurements of the frequency profile. We also used the frequency derivative to evaluate the rate at which the centroid frequency changes on certain timescales. 

\section[]{RESULTS}
\label{res}
\begin{figure}
\begin{center}
\includegraphics[width=80mm]{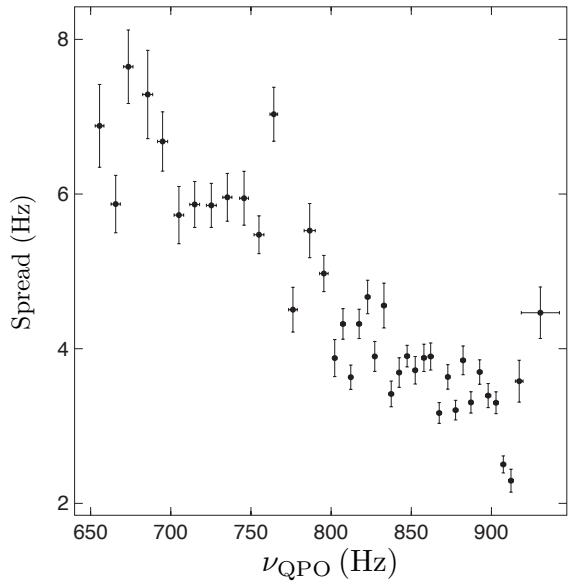}
\caption{Measurements of the frequency spread of the lower kHz QPO in 4U 1636--53 as a function of the QPO centroid frequency. Each point is the average of the QPO-frequency spread at that frequency, corresponding to timescales between 320 s and 1600 s. See section \ref{res} for more details.}
\label{spread}
\end{center}
\end{figure}

In Figure \ref{spread} we show the spread of the lower kHz QPO frequency as a function of the QPO frequency measured over timescales of 320 and 1600 s.  Each point represents the average spread within frequency bins of 5 to 40 Hz, depending on the number of measurements at different frequencies. Notice that the rebin has been applied under the assumption that the spread only depends on frequency.

Figure \ref{spread} shows that the frequency spread decreases from $\sim$ 7 Hz to $\sim$ 2 Hz as the QPO frequency increases from $\sim$ 650 Hz up to $\sim 910$ Hz. Above $\sim 910$ Hz, the data show a marginal increase of the spread to $\sim$ 4 Hz as the frequency increases further up to $\sim$ 930 Hz. The apparent rise of the spread at high frequency is suggested by two points only, which are the average of 13 and 12 measurements, respectively. We fitted the data both with a line and a broken line function, and we carried out an F-test to compare both fits. The F-test probability is $2.2 \times 10^{-2}$, indicating that the fit with a broken line is marginally better than a fit with line. We cannot rule out the possibility that this increase is real, but as we discuss below, it could also be (partly) due to a larger contribution of the statistical errors to the QPO frequency in that frequency range.

To take into account the contribution of the statistical errors from the measurements of the frequency on the spread, we subtracted in quadrature the 1-sigma errors from the observed spread (see \citealt{vau2003}, section 6.1, Eq. 8). We found that the errors of the QPO frequency contribute about 10\% of the observed spread at around 650 Hz, whereas this contribution decreases down to 5\% at 850 Hz, and then increases up to 40\% at $\sim$ 930 Hz. Although this correction slightly flattens the relation, the overall trend does not change significantly, except that the apparent increase of the spread at the high-frequency end becomes less significant.

The accuracy with which one measures the centroid frequency of the QPO depends on the amplitude and the width of the oscillation (e.g., \citealt{v1997}), which in turn change with QPO frequency (e.g., \citealt*{bom2006}). In order to verify whether this affects the relation between spread and frequency, we made simulations where we added normally distributed noise to the frequency profile, with standard deviation ranging from 1 to 10 Hz. As expected, the extra noise component tends to flatten the relation, but errors of up to 10 Hz (rms) in the determination of the QPO frequency do not significantly change the trend shown in Figure \ref{spread}.

\begin{figure}
\begin{center}
\includegraphics[width=80mm]{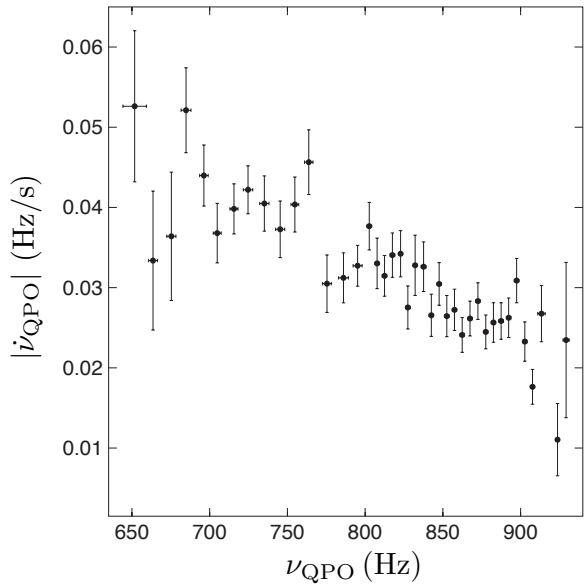}
\caption{Absolute value of the frequency derivative of the lower kHz QPO frequency in 4U 1636--53 as a function of the QPO frequency. Each point is the average value (within a frequency bin) of the frequency derivative of the QPO frequency detected on timescales of 64 s or less. See section \ref{sp} for more details.}
\label{speed64}
\end{center}
\end{figure}

Figure \ref{speed64} shows the absolute value of the frequency derivative of the lower kHz QPO as a function of the QPO frequency. This plot is based on measurements of the QPO frequency on timescales of 64 s or less. Each point represents the average value of the frequency derivative within a frequency bin from 5 to 30 Hz. Similar to what happens with the spread, the frequency derivative decreases as the frequency of the QPO increases, but unlike Figure \ref{spread}, in this case there is no indication of an increase of the frequency derivative above $\sim$ 910 Hz.

\begin{figure}
\begin{center}
\includegraphics[width=80mm]{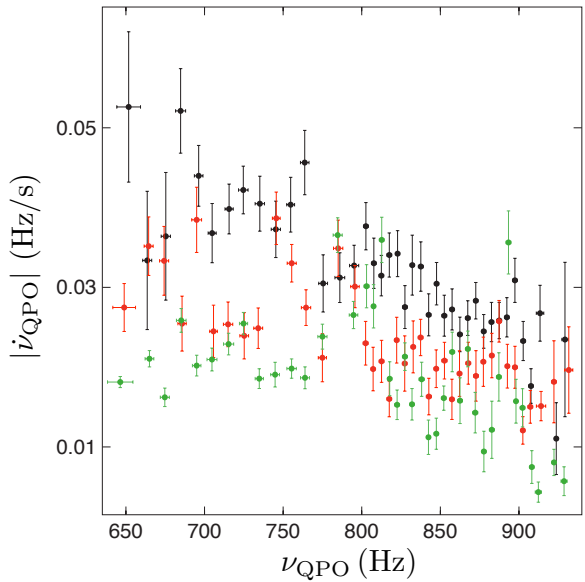}
\caption{Absolute frequency derivative values of the lower kHz QPO in 4U 1636--53 for 3 different timescales. Black points are measurements for QPOs detected on intervals shorter than 64 s. Red points represent detections between 64 s and 160 s, while green points represent measurements of the QPO on intervals of 160 seconds.}
\label{speed_allT}
\end{center}
\end{figure}

In Figure \ref{speed_allT} we compare the absolute value of the frequency derivative of the lower kHz QPOs detected over three different timescales. Black points represent the frequency derivative of the QPOs measured on timescales of 64 s or less (the same as Figure \ref{speed64}), red points are for timescales between 64 and 160 s, and green points are for timescales of 160 s. Interestingly, for timescales longer than 64 s the frequency derivative initially stays more or less constant as the QPO frequency increases, and at around 800 Hz it decreases following more or less the same trend seen on the shortest timescale. We investigated whether the difference between trends in Figure \ref{speed_allT} is only caused by the different timescales. To do that we took the observations where we detected kHz QPOs in segments of data shorter than 64 s (black points in Figure \ref{speed_allT}), we created PDS increasing the length of the data segments up to 160 s and we searched again for kHz QPOs. We then recalculated the frequency derivative and we compared it with the green points in Figure \ref{speed_allT}. We found that the two distributions of points are consistent with each other, which shows that the difference between the three groups of points in Figure \ref{speed_allT} is likely due to the different timescales over which we measured the QPO frequency.

\begin{figure}
\begin{center}
\includegraphics[width=80mm]{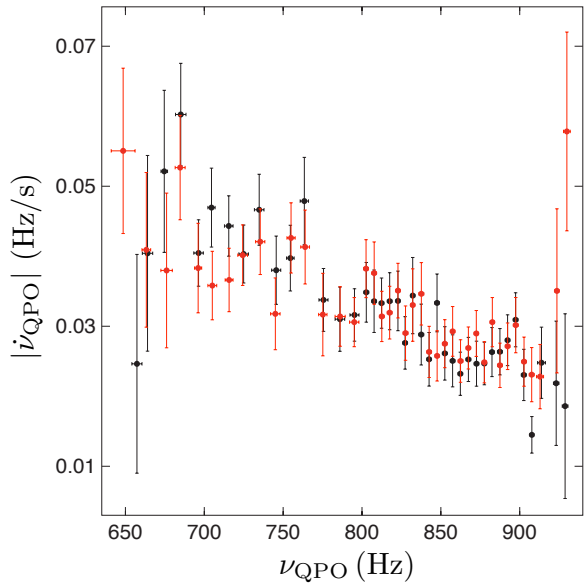}
\caption{Comparison between absolute values of the positive (red points) and negative frequency derivative (black points) for the lower kHz QPOs in 4U 1636--53 detected on timescales shorter than 64 s.}
\label{speed_pos_neg}
\end{center}
\end{figure}

Finally, in Figure \ref{speed_pos_neg} we show, respectively, the absolute value of the negative (frequency decrease) and positive (frequency increase) frequency derivative for kHz QPOs detected on timescales of up to 64 s. We fitted both relations with a linear function. We found that, within the errors, there is no significant difference between the trend of the positive and negative QPO-frequency derivative with QPO frequency.

\section{Discussion}

We detected kHz QPOs in 528 out of 1280 RXTE observations of the LMXB 4U 1636--53 (all RXTE data available of this source up to May 2010). 26 of these 528 observations showed simultaneously the lower and the upper kHz QPO, whereas in the remaining 502 observations we detected a single kHz QPO that we identified using the frequency vs. hard colour diagram (see section \ref{QPOiden}). Using measurements of the lower kHz QPO on timescales between 32 and 160 s we measured, for the first time, the rate at which the QPO frequency changes as a function of the QPO frequency itself. For this we used both the spread of the QPO frequency over groups of 10 consecutive measurements, sampling timescales between 320 and 1600 s, and the time derivative of the QPO frequency over timescales of 32 to 160 s. Both the QPO-frequency spread and QPO-frequency derivative decrease by a factor of $\sim$ 3 as the QPO centroid frequency increases from $\sim$ 600 to $\sim$ 900 Hz. Above $\sim$ 910 Hz the QPO-frequency spread appears to increase again, albeit the evidence for this is marginal. We found that the relation between QPO frequency, $\nu_{\rm QPO}$, and QPO-frequency derivative, $\dot{\nu}_{\rm QPO}$, depends upon the timescale over which the QPO frequency is measured: When $\nu_{\rm QPO}$ is between 600 and $\sim$ 770 Hz, $\dot{\nu}_{\rm QPO}$ decreases by a factor of $\sim$ 2 as the timescale over which we measure the frequency derivative increases from 64 s or less up to 160 s. Finally, we found that the positive and negative values of $\dot \nu_{\rm QPO}$ are consistently distributed in the whole frequency range (see Figure \ref{speed_pos_neg}), even at high frequencies where the accretion disc should approach the ISCO. 

\citet*{bom2006} found that in 4U 1636--53, the coherence of the lower kHz QPO increases from $\sim$10 to $\sim$220 as the frequency of the QPO increases from $\sim$600 to $\sim$850 Hz, and then it decreases rather abruptly as the frequency continues increasing up to 950 Hz\footnote{Data in Figure \ref{spread} to \ref{speed_pos_neg} show no unusual change in the trend of the QPO frequency spread or the time derivative at around 850 Hz, where \citet*{bom2006} see the sudden drop of the coherence of the lower kHz QPO in 4U 1636--53. Similarly, there are no unusual features at frequencies corresponding to small integer ratios of the kHz QPO frequencies in the spread and $\dot \nu_{\rm QPO}$ relations (\citealt{a2003}; \citealt{bmh2005}).}. \citet*{bom2006} were able to model the behaviour of the coherence of the kHz QPO in terms of the lifetime of blobs of matter that move in quasi-Keplerian orbits at the inner edge of a geometrically-thin disc, the width of the disc annulus where these blobs produce the kHz QPOs, and the advection speed of the gas in the disc. \citet*{bom2006} conclude that the abrupt drop of the QPO coherence at about 850 Hz is due to the presence of the ISCO around the neutron star in this and other systems.

Comparing our results with Figure 8 in \citet{bom2006}, we find that the variation of the QPO frequency on timescales between 320 and 1600 seconds contributes significantly to the QPO width in the frequency range between $\sim$750 and $\sim$870 Hz, whereas it does not contribute to the increase of the QPO width at higher QPO frequencies. Previous results show that variations on shorter timescales can have a large impact on the width of QPO. E.g., \citet{ykj2001}, showed that in the neutron star system Sco X-1 the frequency of the upper kHz QPO changes by ~20 Hz on timescales of ~1/6 Hz, which they relate to flux variation on these timescales.

The trend of the QPO spread in Figure \ref{spread} shows a marginal increase of the spread at around 915 Hz, which is close to the frequency at which \citet*{b2005c} identify the signature of the ISCO on the lower kHz QPO in 4U 1636--53. If this apparent rise was real, it could be related to the idea that the high-frequency drop of the QPO coherence is a signature of the disc reaching the ISCO.

Most of the models so far proposed assume that the kHz QPOs are produced in an optically-thick geometrically-thin accretion disc, and that the frequency of either the lower or the upper kHz QPO represents the Keplerian frequency at a given radius in the disc (e.g., \citealt*{mlp1998}; \citealt{sv1998};\citealt{ot1999}). For instance, in the sonic-point model  \citep*{mlp1998} the frequency of the upper kHz QPO corresponds to the radius at which the gas starts spiralling inward at a supersonic radial velocity, and changes of the QPO frequency reflect changes of mass accretion rate (ram pressure and radiation drag). In other models (e.g., \citealt{sv1998}; \citealt{ot1999}) the frequency of the QPOs also depends upon the radius of the disc where the QPOs are produced, although the mechanism by which this radius changes is not specified. In those models mass accretion rate is also likely responsible for changes of the radius in the disc where the QPO is produced.

\subsection{Comparison with standard disc theory}

We can compare our measurements of $\dot{\nu}_{\rm QPO}$ of the lower kHz QPO with what is expected in the case of a thin disc \citep{ss1973} if $\nu_{\rm QPO}$ is the Keplerian frequency very close to the inner edge of the disc. We will further assume that the disc is truncated by an unspecified mechanism. According to the standard disc model \citep{ss1973}, matter in the disc moves radially with a speed given by:
\begin{equation}
v_{r}=0.98\, \alpha_{s}^{4/5}\, \dot{M}^{3/10}\, M_{\star}^{-1/4}\,R^{-1/4}\,f^{-14/5}\,\text{km/s},\\
\end{equation}
with $f=[1-(R_{\star}/R)^{1/2}]^{1/4}$, where $\alpha=0.1\alpha_{s}$ is the viscosity parameter, $\dot{M}$ is the mass accretion rate in units of $2.6\times10^{17}$ g/s, $M_{\star}=M/1.8M_{\odot}$, $R$ is the radius of the inner edge of the disc in km, and $R_{\star}$ is the neutron star radius in units of 14 km, respectively\footnote{We assumed an average luminosity value for 4U 1636--53 of about 0.1 $L_{\rm Edd}$ \citep{b3} for a distance of 5.5 Kpc \citep{vpw1995}, where $L_{\rm Edd}$ is the Eddington luminosity for a 1.8-solar masses neutron star.}. This is the speed at which matter very close to the inner edge of the accretion disc will move inwards one the disc has been truncated. Combining this expression with the Keplerian formula, we find that the derivative of the Keplerian frequency can be expressed as:\\
\begin{equation}
\label{nu_dot}
|\dot{\nu}| \approx 300\,\alpha_{s}^{4/5}\, \dot{M}^{3/10}\, M_{\star}^{-2/3}\,\nu_{kHz}^{11/6}\,F(\nu_{kHz})\,\text{Hz/s},
\end{equation}
where $\nu_{kHz}$ is the Keplerian frequency in units of 1000 Hz, and $F(\nu_{kHz})$ is the factor $f^{-14/5}$ expressed as a function of $\nu_{kHz}$, and normalised by the same factor calculated for a frequency of 1000 Hz. Assuming a 1.8 $M_{\sun}$ neutron star, the function $F(\nu_{kHz})$ increases from $\sim0.6$ to $\sim0.9$ as the frequency increases from 650 to 930 Hz.
From this equation it is apparent that the time derivative of the Keplerian frequency very close to the inner edge of a thin disc increases as the Keplerian frequency increases, contrary to what we observe in the case of the lower kHz QPO in 4U 1636--53. Not only that, but for nominal values of the parameters, the time derivative of the Keplerian frequency in the frequency range between 650 and 950 Hz, is about 4 orders of magnitude larger than what we observe for the lower kHz QPO in 4U 1636--53. As it is apparent from eq. \eqref{nu_dot}, this difference cannot be accounted for by changing the mass and the radius of the NS.

One way to reconcile the observations with the expectations from a thin disc is if the viscosity parameter $\alpha$ is $\sim10^{-5}$ (cf. \citealt{b1997}), although this would not revert the trend of $\dot \nu_{\rm QPO}$ with the QPO frequency. The $\dot \nu_{\rm QPO}$ vs. $\nu_{\rm QPO}$ trend from our analysis could be in principle reproduced by eq.\eqref{nu_dot} if the product $\alpha \times \dot{M}$ depended upon the accretion-disc radius. If we assume for simplicity that $\alpha$ is constant with radius, the trend in eq.\eqref{nu_dot} would be reversed if the mass accreted from the secondary flowed out of the disc at a rate that increases as the radius decreases. In order to fit the observations, in 4U 1636--53 the amount of accreted mass that crosses through the inner edge of the disc must be $\sim 5$\,\% of the total accreted mass from the secondary. In other words, about 95\,\% of the mass accreted from the secondary should not flow onto the neutron star via de accretion disc. Although this scenario cannot be completely ruled out, it is very unlikely because such a process should leave some trace in the spectral properties of the system, e.g., extra low-energy absorption, that has not been observed so far.

The majority of the models of the kHz QPOs propose that it is the {\em upper} kHz QPO that is Keplerian (e.g., \citealt*{mlp1998}, \citealt{sv1998}). Since the upper kHz QPO in 4U 1636--53 is much broader and weaker than the lower one (e.g.; \citealt*{b2005b}), the upper kHz QPO is in general less significant than the lower one when both are measured over the same timescale. We therefore were unable to recover the frequency profile of the upper kHz QPO on the same timescales as for the lower kHz QPO to compare the time derivative of the upper kHz QPO with that of the Keplerian frequency in a thin disc. However, \citet{bmh2005} showed that in 4U 1636--53 the frequency of the lower and upper kHz QPOs, $\nu_l$ and $\nu_u$, respectively, follow a linear relation $\nu_u = 0.673\times\nu_l + 539$ Hz. If this relation still holds on short timescales, we can combine it with our measurements of the lower kHz QPO and estimate the time derivative of the upper kHz QPO frequency. We find that, similar to the case of the lower kHz QPO described above, the time derivative of the frequency of the upper kHz QPO decreases as the frequency of the upper kHz QPO increases, which is again at variance with what is expected from the thin disc model.

From the above we conclude that changes of the QPO frequency, at least for the case of 4U 1636--53, are not compatible with the dynamics of the inner edge of a standard $\alpha$-disc.

\subsection{Standard disc theory plus radiation drag}
\label{miller}
The positive and negative derivatives of the QPO frequency appear to follow the same trend as a function of the QPO frequency (see section \ref{res} and Figure \ref{speed_pos_neg}). This suggests that there is a coupling between the mechanisms that drive the inner edge of the disc inwards and outwards. In this subsection we explore the case in which the accretion disc is truncated at the so-called sonic-point radius, the position in the disc where the radial velocity of the inflowing gas changes from subsonic to supersonic \citep*{mlp1998}. In this scenario, mass accretion rate through the accretion disc pushes the inner edge of the disc inwards in a similar way to the one we described in the previous subsection; but in this case, the radiation produced by the mass that eventually accretes onto the neutron-star surface removes angular momentum from the disc. The position of the inner radius of the accretion disc is therefore set by the interplay between radiation drag and viscosity. Using general relativistic calculations of the gas dynamics and radiation transport in the inner edge of the accretion disc, \citet*{mlp1998} found that the sonic radius can be written approximately as:
\begin{equation}
\label{radius_mill}
R_{\textsl{sr}} \approx R+5\left(\frac{\dot{M_i}}{0.01\dot{M_E}}\right)^{-1} \left(\frac{R}{10\text{km}}\right)\left(\frac{h/r}{0.1}\right)\left(\frac{v^{r}}{10^{-5}\textsl{c}}\right),
\end{equation}
where $R$ is the NS radius in kilometres, $\dot M_i$ is the mass accretion rate through the disc, $\dot M_{E}$ is the Eddington mass accretion rate, $h$ is the thickness of the disc at a radial distance $r$ in the disc, and $v^r$ is the radial velocity of the gas in the accretion disc. We can therefore write:
\begin{equation}
|\dot{\nu}_{\textsl{sr}}| = \frac{3}{4\pi} \left(\textsl{GM}\right)^{1/2}\,\textsl{R}^{-5/2}_{\textsl{sr}}\,|\dot{\textsl{R}}_{\textsl{sr}}|,
\end{equation}
with
\begin{equation}
\label{nu_sr}
|\dot{\textsl{R}}_{\textsl{sr}}|\approx 0.05\,\dot{M}_{E}\left(\frac{\dot{M_i}}{0.01\dot{M}_{E}}\right)^{-2}\ddot{M_i} \left(\frac{R}{10\text{km}}\right)\left(\frac{h/r}{0.1}\right)\left(\frac{v^{r}}{10^{-5}\text{c}}\right),
\end{equation}
where $\ddot M_i$ is the time derivative of the mass accretion rate through the disc.
We can now in principle use equations (3) to (5) to calculate $\dot \nu_{\rm sr}$ vs. $\nu_{\rm sr}$, given the NS mass and radius. For that we also need to specify $h/r$, $v^{r}$ and $\ddot M_i$. In systems in which accretion is the main source of radiation, the bolometric luminosity is proportional to the total mass accretion rate. To the extent that in LMXBs the X-ray intensity, $I_{\rm X}$, is a good measure of the bolometric luminosity (e.g., \citealt{b3}), $\ddot M_i \propto \dot I_{\rm X}$. Putting all this together, and assuming that $h/r$ and $v^r$ are constant, we find that $\dot \nu_{\rm sr} \propto \nu_{\rm sr}^{5/3}\, \dot I_X / I_X^2$. Using the observed variations of $I_{\rm X}$ with time we find that $\dot\nu_{\rm sr}$ decreases by a factor $\sim 2$ as $\nu_{\rm sr}$ increases from $\sim 650$ to $\sim 900$ Hz, which is comparable to what we found in 4U 1636--53 (see Figure \ref{speed64}).

\begin{figure}
\begin{center}
\includegraphics[width=80mm]{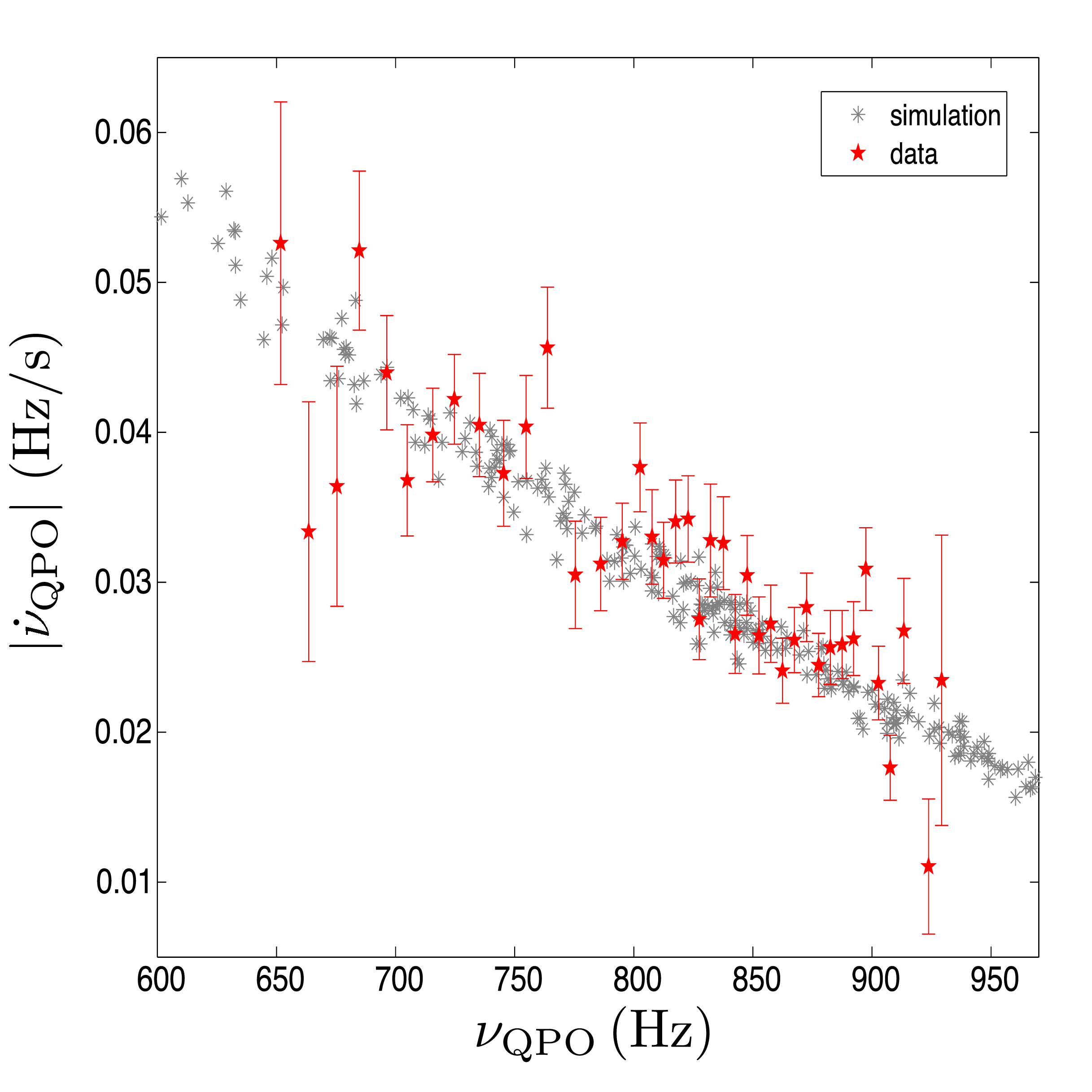}
\caption{Comparison between the absolute value of the frequency derivative of the lower kHz QPO frequency in 4U 1636--53 (red points) and simulated values of the frequency derivative of the lower kHz QPO using the \emph{sonic-point model} from \citet*{mlp1998} (grey stars). See section \ref{miller} for more details about the simulation.}
\label{simultion_vs_data}
\end{center}
\end{figure}

However, it is well known that, while on timescales of hours kHz QPO frequency correlates with $I_{\rm X}$, on timescales of days or longer the kHz QPO frequency can be the same even if $I_{\rm X}$ is a factor of $\sim 2$ different. This is the so-called parallel-track phenomenon (\citealt*{m1999}, \citealt*{v2001}). If the frequency of (one of) the kHz QPOs reflects the Keplerian frequency at the inner edge of the disc, and the position of the inner edge of the disc depends on $\dot M_i$ as in eq. \eqref{radius_mill}, the parallel-track phenomenon means that there cannot be a one-to-one relation between $\dot M_i$ and $I_{\rm X}$, and that the procedure that we described in the previous paragraph is incorrect. To overcome this problem, we proceeded as follows: Since the frequencies of the kHz QPOs appear to follow a random walk (\citealt{bmh2005}; \citealt{bv2012}), we generated values of $\dot M_i$ following a random walk\footnote{We restricted $\dot{M_i}$ to be between 1\% and 3\% $\dot{M}_E$ generating uniformly random distributed steps between $-10^{-4}$ and $10^{-4}$ $\dot{M}_E$.} we calculated $\ddot M_i$, and we computed $\dot \nu_{\rm sr}$ vs. $\nu_{\rm sr}$ using eq. (3) to (5). Following \citet*{mlp1998}, we took $h/r = 0.1$ and $v^r = 10^{-5} c$, we assumed a NS with a radius of 13 km and a mass of 1.9 $M_\odot$, and generated values of $\dot M_i$ between 0.01 and 0.03 $\dot M_E$. We further assumed that the Keplerian frequency at the sonic-point radius is the upper kHz QPO (\citealt*{mlp1998}; but see below), and converted the sonic-point Keplerian frequency to that of the lower kHz QPO using the relation $\nu_u = 0.673 \times \nu_l + 539$ Hz from  \citet{bmh2005}, where $\nu_l$ and $\nu_u$ are, respectively, the frequencies of the lower and the upper kHz QPO. In Figure \ref{simultion_vs_data} we plot the observed relation $\dot \nu_{\rm QPO}$ vs. $\nu_{\rm QPO}$ (red points; these are the same data as in Figure \ref{speed64}) together with the relation between $\dot \nu_{\rm sr}$ and $\nu_{\rm sr}$ from this simulation (grey stars). From this Figure it is apparent that the simulation reproduces the observed data well. We note that although the range of $\dot M_i$ in our simulation is a factor of $\sim 3$ less than the range of luminosities spanned by 4U 1636--53 (e.g., \citealt{b3}), this may be accounted for if we adjust the values of $h/r$ and $v^r$ accordingly. Finally, we note that we can reproduce the data equally well if we assume that the frequency at the sonic-point radius is the lower kHz QPO if, in this case, the neutron star has a radius of 14 km and a mass of 1.7 $M_\odot$.

In summary, assuming that the frequency of one of the kHz QPOs is equal to the Keplerian frequency at the inner edge of the accretion disc, the comparison between the observations and our simulations appears to lend support to the idea that the inner edge of the accretion disc is set by the interplay between viscosity and radiation drag. Our results, however, do not shed light about the mechanism that produces the kHz QPOs, or whether it is the lower or the upper kHz QPO the one that reflects the Keplerian frequency at the inner edge of the accretion disc. It remains to be seen whether other mechanisms (e.g., magnetic drag) can also reproduce the observations.

\section*{Acknowledgments}
This research has made use of data obtained from the High Energy Astrophysics Science Archive Research Center (HEASARC), provided by NASA's Goddard Space Flight Center. This research made use of the SIM-BAD database, operated at CDS, Strasbourg, France and NASA's Astrophysics Data System. We thank the International Space Science Institute (ISSI) for their support. We thank Cole Miller, Didier Barret, Fred Lamb, Jeroen Homan, Phil Uttley and Luciano Burderi for interesting discussions on the topic. We are grateful to an anonymous referee for the comments that helped us improve the manuscript. TMB acknowledges support from grant INAF-ASI 1/009/10/0.


\begin{thebibliography}{99}
\bibitem[\protect\citeauthoryear{Abramowicz et al.}{2003}]{a2003}Abramowicz M.~A., Karas V., Kluzniak W., Lee W.~H., Rebusco P., 2003, PASJ, 55, 467
\bibitem[\protect\citeauthoryear{Altamirano et al.}{2008}]{a2008} Altamirano D., van der Klis M., M\'endez M., Wijnands R., Markwardt C., Swank J., 2008, ApJ, 687, 488
\bibitem[\protect\citeauthoryear{Barret et al.}{2005a}]{b2005a}Barret D., Klu\'zniak W., Olive J.~F., Paltani S., Skinner G.~K., 2005, MNRAS, 357,1288
\bibitem[\protect\citeauthoryear{Barret et al.}{2005b}]{b2005b}Barret D., Olive J.-F., Miller M.~C., 2005, MNRAS, 361, 855
\bibitem[\protect\citeauthoryear{Barret et al.}{Barret, Olive \& Miller}{2005c}]{b2005c}Barret D., Olive J.-F., Miller M.~C., 2005, AN, 326, 808
\bibitem[\protect\citeauthoryear{Barret et al.}{Barret, Olive \& Miller}{2006}]{bom2006}Barret D., Olive J.-F., Miller M.~C., 2006, MNRAS, 370, 1140
\bibitem[\protect\citeauthoryear{Barret \& Vaughan}{2012}]{bv2012}Barret D.,Vaughan S., 2012, ApJ, 746, 131
\bibitem[\protect\citeauthoryear{Belloni et al.}{1997}]{b1997}Belloni T., M\'endez M., King A. R., van der Klis M., van Paradijs, 1997, ApJ, 488, L109
\bibitem[\protect\citeauthoryear{Belloni, Mendez \& Homan}{Belloni et al.}{2005}]{bmh2005} Belloni T., Mendez M., Homan J., 2005, A\&A, 437, 209
\bibitem[\protect\citeauthoryear{Belloni et al.}{2007}]{b2007}Belloni T., Homan J., Motta S., Ratti E., M\'endez, M., 2007, MNRAS, 379, 247
\bibitem[\protect\citeauthoryear{Berger et al.}{1996}]{b1996}Berger M., van der Klis M., van Paradijs J., Lewin W.~H.~G., Lamb F., Vaughan B., Kuulkers E., Augusteijn T., Zhang W., Marshall F.~E., Swank J.~H., Lapidus I., Lochner J.~C., Strohmayer T.~E., 1996, ApJ, 469L, 13
\bibitem[\protect\citeauthoryear{Bradt, Rothschild \& Swank}{1993}]{b1993}Bradt H.~V., Rothschild R.~E., Swank J.~H., 1993, A\&AS, 97, 355
\bibitem[\protect\citeauthoryear{Di Salvo et al.}{2001}]{ds2001} Di Salvo T., M\'endez M., van der Klis M., Ford E., Robba N.~R., 2001, ApJ, 546, 1107
\bibitem[\protect\citeauthoryear{Di Salvo et al.}{Di Salvo, M\'endez and van der Klis}{2003}]{dmv2003}Di Salvo T., M\'endez M., van der Klis M., 2003, A\&A, 406, 177
\bibitem[\protect\citeauthoryear{Ford et al.}{2000}]{b3} Ford E.~C., van der Klis M., M\'endez M., Wijnands R., Homan J., Jonker P.~G., van Paradijs J.,2000 , ApJ, 537, 368
\bibitem[\protect\citeauthoryear{Gierli\'nski \& Done}{2002}]{gd2002}Gierli\'nski M., Done C., 2002, MNRAS, 337,1373G 
\bibitem[\protect\citeauthoryear{Homan et al.}{2002}]{h2002}Homan J., van der Klis M., Jonker P.~G., Wijnands R., Kuulkers E., M\'endez M., Lewin W.~H.~G., 2002, ApJ, 568, 878
\bibitem[\protect\citeauthoryear{Jahoda et al.}{2006}]{ja2006}Jahoda K., Markwardt C.~B., Radeva Y., Rots A.~H., Stark M.~J., Swank J.~H., Strohmayer T.~E., Zhang W., 2006, ApJS, 163, 401
\bibitem[\protect\citeauthoryear{Jonker et al.}{2000}]{j2000}Jonker P.~G., van der Klis M., Wijnands R., Homan J., van Paradijs J., M\'endez M., Ford E.~C., Kuulkers E., Lamb F.~K., 2000, ApJ, 537, 374
\bibitem[\protect\citeauthoryear{Kuulkers et al.}{1994}]{k2004} Kuulkers E., van der Klis M., Oosterbroek T., Asai K., Dotani T., van Paradijs J., Lewin W.~H.~G., 1994, A\&A, 289, 795
\bibitem[\protect\citeauthoryear{Lin, Remillard \& Homan}{2007}]{lrh2007}Lin D., Remillard R., Homan J., 2007, ApJ, 667, 1073
\bibitem[\protect\citeauthoryear{Matt et al.}{1991}]{m1991}Matt G., Perola G. \& Piro L., 1991, A\&A, 247, 25M
\bibitem[\protect\citeauthoryear{M\'endez et al.}{1999}]{m1999} M\'endez M., van der Klis M., Ford E.~C., Wijnands R., van Paradijs J., 1999, ApJ, 511L, 49
\bibitem[\protect\citeauthoryear{M\'endez and van der Klis}{1999}]{mv1999} M\'endez M., van der Klis M., 1999, ApJ, 517L, 51
\bibitem[\protect\citeauthoryear{M\'endez et al.}{M\'endez, van der Klis \& Ford}{2001}]{mvf2001}M\'endez M., van der Klis M., Ford E.~C., 2001, ApJ, 561, 1016
\bibitem[\protect\citeauthoryear{M\'endez}{2006}]{m2006} M\'endez M., 2006, MNRAS, 371, 1925
\bibitem[\protect\citeauthoryear{Miller et al.}{Miller, Lamb \& Psaltis}{1998}]{mlp1998} Miller M.~C., Lamb F.~K, Psaltis D., 1998, ApJ, 508, 791
\bibitem[\protect\citeauthoryear{Narayan \& Yi}{1995}]{ny1995}Narayan R., Yi I., 1995, ApJ, 452, 710
\bibitem[\protect\citeauthoryear{Press \& Vetterling}{1989}]{nr}Press S.A.T. Brian P.~Flannery, Vetterling W.T., 1989, Numerical Recipes (Fortran Version), Cambridge University Press
\bibitem[\protect\citeauthoryear{Sanna et al.}{2010}]{sa2010} Sanna A., M\'endez M., Altamirano D., Homan J., Casella P., Belloni T., Lin D., van der Klis M., Wijnands R.,  2010, MNRAS, 408, 622
\bibitem[\protect\citeauthoryear{Savitzky \& Golay}{1964}]{sk1964} Savitzky, A., \& Golay, M.~J.~E.\ 1964, Analytical Chemistry, 36, 1627
\bibitem[\protect\citeauthoryear{Shakura \& Sunyaev}{1973}]{ss1973}Shakura, N. I., Sunyaev, R. A, 1973, A\&A, 24, 337-355
\bibitem[\protect\citeauthoryear{Stella \& Vetri}{1998}]{sv1998} Stella L., Vetri M., 1998, ApJ, 492, L59
\bibitem[\protect\citeauthoryear{Strohmayer et al.}{1996}]{st1996}Strohmayer T., Zhang W., Swank J., 1996, IAUC, 6320
\bibitem[\protect\citeauthoryear{ Osherovich \& Titarchuk}{1999}]{ot1999} Osherovich, V. , Titarchuk, L., 1999, ApJ, 522, 1130
\bibitem[\protect\citeauthoryear{van der Klis et al.}{1996a}]{v1996a}van der Klis M., Swank J.~H., Zhang W., Jahoda K., Morgan E.~H., Lewin W.~H.~G., Vaughan B., van Paradijs J. , 1996, IAUC, 6319
\bibitem[\protect\citeauthoryear{van der Klis et al.}{1996b}]{v1996} van der Klis M., Swank J.~H., Zhang W., Jahoda K., Morgan E.~H., Lewin W.~H.~G., Vaughan B., van Paradijs J. , 1996, ApJ, 469
\bibitem[\protect\citeauthoryear{van der Klis}{1997}]{v1997}van der Klis M., 1997, ASSL, 218, 121
\bibitem[\protect\citeauthoryear{van der Klis}{2001}]{v2001}van der Klis M., 2001, ApJ, 561, 943
\bibitem[\protect\citeauthoryear{van der Klis}{2005}]{v2005}van der Klis M., 2005, AN, 326, 798
\bibitem[\protect\citeauthoryear{van Paradijs \& White}{1995}]{vpw1995}van Paradijs J., White N., 1995, ApJ, 447, L33
\bibitem[\protect\citeauthoryear{van Straaten et al.}{2000}]{van2000} van Straaten S., Ford E.~C., van der Klis M., M\'endez M., Kaaret, P., 2000, ApJ, 540, 1049
\bibitem[\protect\citeauthoryear{van Straaten et al.}{2002}]{van2002} van Straaten S., van der Klis M., di Salvo T., Belloni T., 2002, ApJ, 568, 912
\bibitem[\protect\citeauthoryear{van Straaten et al.}{2003}]{van2003} van Straaten S., van der Klis M., M\'endez M., 2003, ApJ, 596, 1155
\bibitem[\protect\citeauthoryear{Vaughan et al.}{2003}]{vau2003} Vaughan S., Edelson R., Warwick R.~S, Uttley P., 2003, MNRAS, 345, 1271
\bibitem[\protect\citeauthoryear{Yu, van der Klis \& Jonker}{2001}]{ykj2001}Yu W., van der Klis M., Kuulkers E., Jonker P., 2001, ApJ, 559, L29-L32
\bibitem[\protect\citeauthoryear{Zhang, M\'endez \& Altamirano}{2011}]{z2011}Zhang G., M\'endez M.,. and Altamirano D., 2011, MNRAS, 413, 1913
\end{thebibliography}
\end{document}